\documentstyle[12pt,epsf]{article}

\newcommand{\postscript}[2]
 {\setlength{\epsfxsize}{#2\hsize}
 \centerline{\epsfbox{#1}}}

\begin{document}

\begin{flushright}
FERMILAB-CONF-99/239-E\\
UPR--0239--E\\
\today\\
\end{flushright}
\bigskip
\begin{center}
{\large CP Violation in B Decays at the Tevatron}\\
\bigskip
I.~Joseph Kroll\\
The University of Pennsylvania\\
David Rittenhouse Laboratory\\
209 South 33rd Street\\
Philadelphia, PA 19104\\
\end{center}

\begin{abstract}
Between 1992 to 1996, the CDF and D0 detectors each collected
data samples exceeding 100~pb$^{-1}$ of $p\bar{p}$ collisions
at $\sqrt{s}=1.8$~TeV at the Fermilab Tevatron. 
These data sets led to a large number of precision
measurements of the properties of $B$~hadrons including lifetimes, masses,
neutral $B$~meson flavor oscillations, and relative branching fractions,
and to the discovery of the $B_c$~meson.
Perhaps the most exciting result was the first look at the
$CP$~violation parameter $\sin(2\beta)$ using the world's
largest sample of fully reconstructed
$B^0/\bar{B}^0\rightarrow J/\psi K^0_{\rm s}$ decays.
A summary of this result is presented here.
In the year 2000, the Tevatron will recommence $p\bar{p}$ collisions
with an over order of magnitude expected increase in integrated luminosity
(1~fb$^{-1}$ per year).
The CDF and D0 detectors will have undergone substantial upgrades,
particularly in the tracking detectors and the triggers.
With these enhancements, the Tevatron $B$ physics program will include
precision measurements of $\sin(2\beta)$ and $B^0_s$ flavor oscillations,
as well as studies of rare $B$~decays that are sensitive to new physics.
The studies of $B^0_s$ mesons will be particularly interesting as this
physics will be unique to the Tevatron during the first half of the next
decade.
\end{abstract}

{\vfill
{\footnotesize{presented at Kaon '99,
Chicago, Illinois, 23 June 1999 \hfill}}}

\newpage

\section{Introduction}

In this paper, we review results on $CP$ violation in $B$ decays
from the Fermilab Tevatron and discuss prospects in the near future.
The results are based on the analysis of 110~pb$^{-1}$
of $p\bar{p}$ collisions at $\sqrt{s}=1.8$~TeV,
collected between August 1992 to February 1996,
which we refer to as Tevatron Run~I.
At present, only the CDF Collaboration has presented results on
$CP$ violation in $B$ decays from these data.

The Tevatron will recommence $p\bar{p}$ collisions in the Spring 2000
at $\sqrt{s}=2.0$~TeV.
This future data taking period is referred to as Run~II.
The new crucial accelerator component, the main injector, has
been commissioned successfully and will increase the rate of production
of antiprotons by a factor of three above previous rates.
The expected data rate is 2~fb$^{-1}$ in the first two years of operation.
This corresponds to approximately $10^{11}$~$b\bar{b}$ pairs per year.

The Tevatron will continue to operate beyond these first two years.
Ultimately a data sample of more than 20~fb$^{-1}$ may be collected
prior to the turn-on of the LHC.
At that time, Fermilab may continue operation of the Tevatron with
an experiment dedicated to the study of $B$~physics.
The physics motivation and capabilities of such a detector are not
discussed here, but an excellent discussion can be found in~\cite{ref:btev}.


As has been discussed in numerous other talks at this conference,
the study of $B$~hadron decays plays a unique role in the test
of the Standard Model of electroweak interactions and
the study of the Cabibbo Kobayashi Maskawa Matrix~$V_{\rm CKM}$.
Measurements of the decays of $B$~hadrons determine the magnitudes
of five of the nine elements of $V_{\rm CKM}$ as well as the phase.

The unitarity of $V_{\rm CKM}$ leads to nine unitarity relationships,
one of which is of particular interest:
\begin{equation}
\label{eqn:UT}
V_{\rm ud}V_{\rm ub}^* + V_{\rm cd}V_{\rm cb}^* + V_{\rm td}V_{\rm tb}^* = 0.
\end{equation}
This sum of three complex numbers forms a triangle in the complex plane,
commonly referred to as {\it the} Unitarity Triangle.
Measurements of the weak decays of $B$~hadrons and the already known CKM
matrix elements determine the magnitudes of the three sides of the Unitarity
Triangle, and $CP$~asymmetries in $B$~meson decays determine the three angles. 
The primary goal of $B$~physics in the next decade is to measure precisely
both the sides and angles of this triangle and test consistency within
the Standard Model.

\begin{figure}[htb]
\postscript{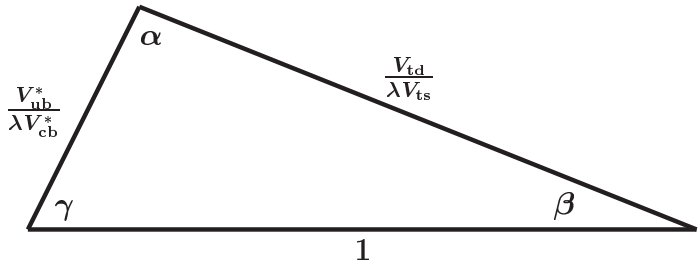}{0.7}
\caption{
A pictorial representation of the Unitary Triangle in the complex plane.
The vertical axis is the complex axis.
%
}
\label{fig:UT}
\end{figure}

We can use several approximations to express Eq.~\ref{eqn:UT}
in a more convenient form.
The elements $V_{\rm ud}\simeq 1$ and
$V_{\rm cd}\simeq -\lambda = -\sin\theta_{\rm C}$, where $\theta_{\rm C}$
is the Cabibbo angle, are well measured.
Although the  elements $V_{\rm tb}$ and $V_{\rm ts}$ are not well
measured, the theoretical expectations are that
$V_{\rm tb} \simeq 1$ and $V_{\rm ts} \simeq -V_{\rm cb}^*$. 
With these assumptions, Eq.~\ref{eqn:UT} becomes
\begin{equation}
\label{eqn:UTREV}
\frac{V_{\rm ub}^*}{\lambda V_{\rm cb}^*}
-1 - \frac{V_{\rm td}}{\lambda V_{\rm ts}} = 0.
\end{equation}
A pictorial representation of this equation as a triangle
in the complex plane is shown in Fig.~\ref{fig:UT};
the vertical axis is the complex axis.
In the Wolfenstein approximation~\cite{ref:wolf,ref:pdg},
the apex of this triangle is $(\rho,\eta)$.

The next experimental steps in studying this triangle will be the
measurement of the angle $\beta$ from the asymmetry in the
decays $B^0/\bar{B}^0\rightarrow J/\psi K^0_{\rm S}$,
and the measurement of $|V_{\rm td}/V_{\rm ts}|$ from the
ratio of mass differences $\Delta m_d/\Delta m_s$ determined
from $B^0 - \bar{B}^0$ and $B^0_s - \bar{B}_s^0$ flavor oscillations.
Since $B^0_s$~mesons are not produced on the $\Upsilon(4S)$~resonance,
the measurement of $B^0_s - \bar{B}_s^0$ flavor oscillations
as well as the study of $B^0_s$ decays will be unique to experiments
operating at hadron machines.


In the $B$ system, the measurements of $CP$~violation that are
cleanly related ({\it i.e.,} without large theoretical uncertainties)
to angles in the Unitarity Triangle are from asymmetries in the
decays of neutral $B$~mesons to $CP$ eigenstates.
The most popular mode is $B^0/\bar{B}^0\rightarrow J/\psi K^0_{\rm S}$.
It is lucky that at hadron colliders the leptonic decays
$J/\psi\rightarrow\mu^+\mu^-$ and $J/\psi\rightarrow e^+e^-$
form the basis of a practical trigger for this decay mode
and the means to reconstruct the decay mode with excellent signal to noise. 

$CP$~violation is observed as an asymmetry ${\cal A}_{CP}$ in decay rates:
%
\begin{equation}
\label{eqn:acp}
{\cal A}_{CP}=\frac
{\frac{dN}{dt}(\bar{B}^0\rightarrow J/\psi\,K^0_{\rm S}) -
 \frac{dN}{dt}(B^0\rightarrow J/\psi\,K^0_{\rm S})}
{\frac{dN}{dt}(\bar{B}^0\rightarrow J/\psi\,K^0_{\rm S}) +
 \frac{dN}{dt}(B^0\rightarrow J/\psi\,K^0_{\rm S})},
\end{equation}
where $\frac{dN}{dt}(\bar{B}^0\rightarrow J/\psi\,K^0_{\rm S})$
is the rate of observed $J/\psi\,K^0_{\rm S}$ given the particle
produced was a $\bar{B}^0$.
%
%
The asymmetry is produced by the interference of direct decays
($\bar{B}^0\rightarrow J/\psi\,K^0_{\rm S}$) and decays that occur after
mixing ($\bar{B}^0 \rightarrow B^0 \rightarrow J/\psi\,K^0_{\rm S}$).
The asymmetry oscillates as a function of proper decay time $t$ with
a frequency $\Delta m_d$, and
the amplitude of the oscillation is $\sin(2\beta)$:
\begin{equation}
\label{eqn:acpt}
{\cal A}_{CP}(t) = \sin(2\beta)\sin(\Delta m_d t).
\end{equation}
Unlike experiments operating at the $\Upsilon(4S)$,
the time integrated asymmetry of Eq.~\ref{eqn:acp}
({\it i.e.} replace the rates in Eq.~\ref{eqn:acp}
with the total observed numbers)
does not vanish in hadron collisions:
\begin{equation}
\label{eqn:acpint}
{\cal A}_{CP}=\frac{x_d}{1+x_d^2}\sin(2\beta) \approx 0.5\sin(2\beta),
\end{equation}
where $x_d=\Delta m_d/\tau(B^0) = 0.732\pm0.032$~\cite{ref:pdg}.
The value of $x_d$ falls fortuitously close to $1.0$, the value that
maximizes the coefficient in front of $\sin(2\beta)$.
In contrast, the large value of $x_s$
($x_s>14.0$ at $95\%$~C.L.~\cite{ref:pdg})
implies that even if there were large $CP$ violation due to $B^0_s$ mixing,
the asymmetry can not be observed using a time integrated asymmetry.
%

Even though the time integrated asymmetry can be used to extract
$\sin(2\beta)$, it is better to measure the asymmetry as a function of
proper decay time ({\it i.e.} the time dependent asymmetry), if possible.
%
%
%
%
The improvement comes from two sources.
First, there is more statistical power in the time dependent asymmetry.
Decays at low lifetime exhibit a small asymmetry
because there has not been adequate time for mixing to occur
to create the interference leading to $CP$ violation.
Second, a substantial fraction of the combinatoric background occurs at
low values of $t$, well below the value of $t$ (about 2.2 lifetimes)
where ${\cal A}_{CP}(t)$ is a maximum.

\section{Current Results}

So far at the Tevatron, only the CDF experiment has had the capability to
measure $CP$~violation in $B$~decays.
This will change in Run~II, when major upgrades of both the CDF and D0
detector will make both detectors capable of unique and important measurements
of $B$~hadron decays.

%
%
%
%
%
%

The features of the Run~I CDF detector~\cite{ref:cdf_runi_det}
crucial for $B$ physics included a four-layer silicon microstrip detector,
a large volume drift chamber, and excellent electron and muon identification.
The silicon microstrip detector provided an
impact parameter ($d_0$) resolution for charged tracks of
$\sigma(d_0) = (13+40/p_T)$\,$\mu$m, where $p_T$ is the
magnitude of the component of the momentum of the track transverse
to the beam line in units of GeV/$c$.
This impact parameter resolution made the precise measurement
of $B$~hadron proper decay times $t$ possible.
The drift chamber was 1.4\,m in radius and was
immersed in a 1.4\,T axial magnetic field.
It provided excellent momentum resolution
$(\delta p_T/p_T)^2=(0.0066)^2 \oplus (0.0009p_T)^2$
(where $p_T$ is in units of GeV/$c$)
and excellent track reconstruction efficiency making
it possible to fully reconstruct $B$~hadron decays
with excellent mass resolution and high signal to noise.
Electron ($e$) and muon ($\mu$) detectors in the central rapidity
region ($|y|<1$) made it possible to detect (and trigger on)
$B$~hadrons using
semileptonic decays ($b\rightarrow\ell X, \ell=e,\mu$) or using
$B\rightarrow J/\psi X, J/\psi\rightarrow \mu^+\mu^-$.

The measurement of ${\cal A}_{CP}(t)$ of Eq.~\ref{eqn:acp}
has three crucial experimental components:
(1) reconstructing the decay mode
$B^0/\bar{B}^0\rightarrow J/\psi K^0_{\rm S}$
with good signal to noise;
(2) measuring the proper decay time $t$; and
(3) determining whether the meson that was produced was
a $B^0$ ({\it i.e.}, $\bar{b}d$) or a $\bar{B}^0$ ({\it i.e.}, $b\bar{d}$).
This last component is known as ``$b$~flavor tagging,''
and it is the most challenging of the above three requirements.
The CDF experiment has demonstrated that the first two are possible
by extracting large-statistics, low-background signals of
fully reconstructed $B$ decays using $J/\psi\rightarrow \mu^+\mu^-$
and partially reconstructed semileptonic decays
({\it e.g.}, $B^0_s\rightarrow\ell^+ D_s^- X$ and charge conjugate).
With these signals, CDF has made some of the most precise species
specific measurements of $B$~hadron lifetimes~\cite{ref:pdg} to date.

The performance of the $b$~flavor tags may be quantified conveniently
by their efficiency~$\epsilon$ and dilution~$D$.
The efficiency is the fraction of $B$ candidates to which the flavor
tag can be applied ($0<\epsilon<1$).
%
%
The dilution is related to the probability ${\cal P}$ that the tag is correct:
$D=2{\cal P} - 1$, so a perfect tag has $D=1$, and a random tag has $D=0$.

The experimentally measured amplitude of the asymmetry in
Eq.~\ref{eqn:acpt} is reduced by the dilution of the tag:
\begin{equation}
\label{eqn:acpmeas}
{\cal A}_{CP}^{\rm meas}(t) = D\sin(2\beta)\sin(\Delta m_d t).
\end{equation}
The statistical error on the true asymmetry ${\cal A}$ is approximately
(for $D^2{\cal A}\ll 1$)
\begin{equation}
\label{eqn:acperr}
\delta {\cal A}\approx\sqrt{\frac{1}{\epsilon D^2 N}},
\end{equation}
where $N$ is the total number of candidates (signal and background)
before applying the flavor tag.
The statistical power of the data sample scales with $\epsilon D^2$.
At a hadron collider, a flavor tag with $\epsilon D^2\ge 1\%$
is respectable.

To extract $\sin(2\beta)$ from the measured asymmetry, the value of
the dilution of the flavor tag(s) must be determined quantitatively.
The most reliable means of determining $D$ is from the data themselves.
Measurements of neutral $B$~meson flavor oscillations requires
$b$~flavor tagging as well.
In this case, the measured asymmetry ${\cal A}_{\rm mix}$ is given by
\begin{equation}
\label{eqn:amix}
{\cal A}_{\rm mix}(t) = 
\frac{N_{\rm unmixed}(t) - N_{\rm mixed}(t)}
{N_{\rm unmixed}(t) + N_{\rm mixed}(t)} =
D\cdot\cos (\Delta m t),
\end{equation}
where $N_{\rm unmixed}(t)$ [$N_{\rm mixed}(t)$] are the number of
candidates observed to have decayed with the same [opposite] $b$~flavor
as they were produced with.
CDF has used measurements of $B^0-\bar{B}^0$ flavor oscillations to
determine $D$ for three different $b$~flavor tagging methods.

The methods of $b$~flavor tagging fall into two categories:
(1) opposite-side flavor tags (OST) and (2) same-side flavor tags (SST).
The dominant production mechanisms of $b$~quarks in hadron collisions
produce $b\bar{b}$ pairs.
Opposite-side flavor tags exploit this fact: to identify the production flavor\
of the $B$~hadron of interest ({\rm e.g.} the one that eventually leads to a
$J/\psi K^0_{\rm S}$), we identify the flavor of the second $B$~hadron
in the event and {\it infer} the flavor at production of the first $B$.
Since the Run~I CDF central drift chamber was fully efficient only
for the central rapidity region ($-1<\eta<1$, where $\eta$ is
pseudorapidity), full reconstruction of $B$ decays and $b$~flavor
tagging were usually restricted to this central region.
If a $J/\psi K^0_{\rm S}$ candidate is detected in this region,
then the second $B$ is in this central region as well only about
$50\%$ of the time.
This means that opposite-side flavor tags at CDF have a maximum
efficiency of $\epsilon=0.5$.

So far, two opposite-side flavor tags have been used:
(1) a lepton tag and (2) a jet-charge tag.
The lepton tag is based on $b$ semileptonic decay: $b\rightarrow\ell^- X$, but
$\bar{b}\rightarrow\ell^+ X$, where $\ell=e,\mu$.
The charge of the lepton identifies the flavor of the $b$.
The low semileptonic branching fraction ($\sim 10\%$ per lepton flavor)
limits the efficiency of this tag, although the flavor tag dilution
is high.
Of course leptons can originate from the decays of secondary decay products
in $B$ decay, {\it e.g.} $b\rightarrow c\rightarrow \ell^+ X$,
and this will cause the wrong production flavor to be assigned,
reducing the dilution.
The jet-charge tag is based on the
momentum weighted charge average $Q^b_{\rm jet}$
of the charged particles produced in the fragmentation
of a $b$ quark and in the subsequent decay of the $B$~hadron.
On average, a $b$~quark will produce a jet charge less than zero
($Q^b_{\rm jet}<0$),
but a $\bar{b}$~quark will produce a jet charge larger than zero
($Q^{\bar{b}}_{\rm jet}>0)$.
The jet-charge tag is more efficient, but has a smaller dilution,
than the lepton-tag. 
The jet-charge tag was especially effective in measurements of
neutral $B$ meson flavor oscillations performed at
$e^+e^-$ colliders operating on the $Z^0$~resonance,
but it is much more challenging to apply this method
in the environment of hadronic collisions.
Both the lepton flavor tag and the jet-charge flavor tag were
used in a precise measurement of $\Delta m_d$~\cite{ref:sltjctprd}
by CDF.

\begin{figure}[htb]
\postscript{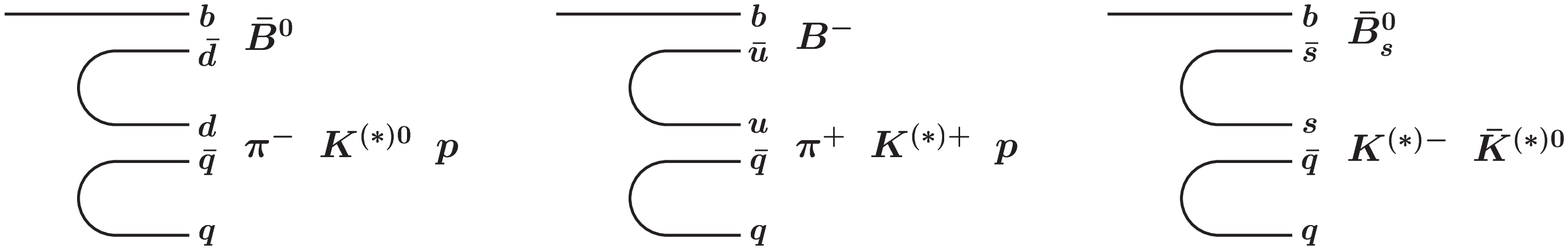}{1.0}
\caption{The same-side flavor tag is based on the correlation
between $b$~flavor and the charge of particles produced in
$b$~quark fragmentation.
}
\label{fig:sst}
\end{figure}

\begin{figure}[htb]
\postscript{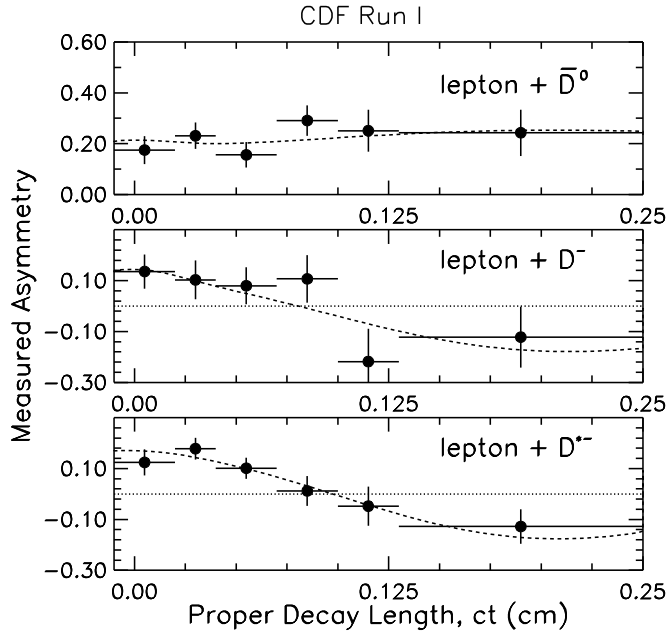}{0.6}
\caption{The mixing asymmetry ${\cal A}_{\rm mix}$ observed in a sample
of partially reconstructed $B^+$ and $B^0$ semileptonic decays using
a same-side $b$~flavor tag.
The plots are described in the text.
}
\label{fig:sstmix}
\end{figure}

The same-side flavor tag exploits the correlation between the
$b$~flavor and the charge of the particles produced by the
$b$~quark fragmentation, as illustrated in Fig.~\ref{fig:sst}.
%
%
A $\pi^+$ tags a $B^0$, but a $\pi^-$ tags a $\bar{B}^0$.
If a charged $B$~hadron is produced, the correlation between
pion charge and $b$~flavor is the {\it opposite} to the correlation
in the case of the $B^0$.
This has important consequences when utilizing fully reconstructed
charged $B$~hadron decays ({\it e.g.}, $B^{\pm}\rightarrow J/\psi K^{\pm}$)
to quantify the performance of the SST when applied to $B^0$.
The idea of the SST was originally proposed by
Gronau, Nippe, and Rosner~\cite{ref:gronau}.
The decays of $P$~wave $B^{**}$~mesons produce the same
$b$ flavor-pion charge correlations.
Since the SST is based on particles produced with the $B$ of interest,
if this $B$ is in the experimental acceptance, then it is likely that
these fragmentation particles are in the acceptance as well.
As a result, the efficiency of the SST is potentially larger than the
efficiency of the opposite-side tags.
The SST was used successfully~\cite{ref:sst} as a flavor tag
in a precise measurement of $\Delta m_d$ by CDF using
approximately 6\,000 partially reconstructed semileptonic
$B^0$ decays in the decay modes $B^0\rightarrow\ell^- D^{(*)+} X$.
The mixing signal is shown in Fig.~\ref{fig:sstmix}.
This figure depicts the experimental asymmetry described by
Eq.~\ref{eqn:amix}.
The points with error bars are the measured asymmetry plotted
as a function of proper decay length $ct$.
%
The upper plot is for $B^+$ decays and the lower two plots
are for $B^0$ decays.
The dashed curves are the expected asymmetry fit to the data.
The two lower curves show the expected $\cos(\Delta m_d t)$ behavior
from $B^0$ mixing.
The intercept on the vertical axis at $ct=0$ is the dilution $D_0$,
where the subscript indicates this is the dilution for neutral $B^0$.
Charged $B^+$ do not oscillate, so the curve in the upper-most
plot is a straight line (the small oscillation is due to a
$\sim 15\%$ contamination of $B^0$ in this sample).
The intercept determines the dilution $D_+$ for the same-side tag
for $B^+$.
The charged dilution $D_+$ is larger than the neutral dilution $D_0$.
Monte Carlo studies~\cite{ref:sst} confirm that this is due to charged kaons,
which increase $D_+$, but decrease $D_0$,
as illustrated in Fig.~\ref{fig:sst}.

\begin{figure}[htb]
\postscript{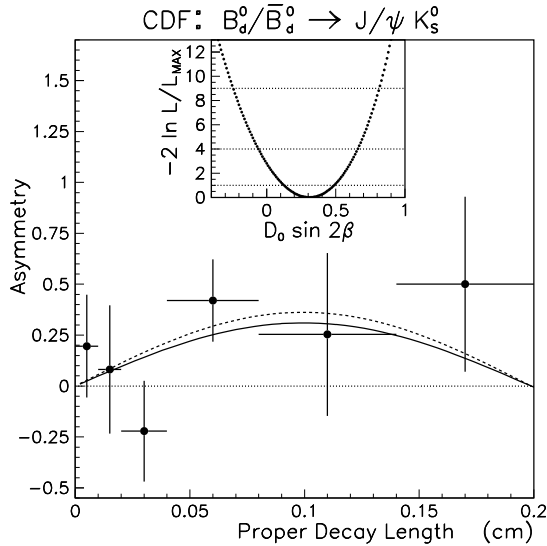}{0.5}
\caption{The measured asymmetry ${\cal A}_{cp}^{\rm meas}$ as a
function of proper decay length $ct$ for
$B^0/\bar{B}^0\rightarrow J/\psi K^0_{\rm S}$ decays.
The plot is described in the text.
}
\label{fig:sin2betaprl}
\end{figure}

Using this same-side flavor tag, and a signal of $198\pm17$
$B^0/\bar{B}^0\rightarrow J/\psi K^0_{\rm S}$ decays,
CDF published~\cite{ref:sin2betaprl}
a first look at $\sin(2\beta)$ from the Run~I data.
The final state is reconstructed using the decays
$J/\psi\rightarrow\mu^+\mu^-$ (this is also the basis of the trigger)
and $K^0_{\rm S}\rightarrow\pi^+\pi^-$.
The measured asymmetry ${\cal A}$ as a function of proper decay length
$ct$ is shown in Fig.~\ref{fig:sin2betaprl}.
The points with error bars are the experimental asymmetries;
the curves are fits of these points to the expected form
given in Eq.~\ref{eqn:acpmeas}.
The resulting amplitude is
$D\sin(2\beta)=0.31\pm0.18\,({\rm stat}.)\pm0.03\,({\rm syst}.)$,
where the largest contributions to the systematic error are
the error on world average value of $\Delta m_d$ and the ability
to constrain potential detector biases that could produce false asymmetries.
If the error on this measurement were small enough to establish
that this amplitude is not zero, then this would be sufficient to
establish $CP$ violation in the $B$ system.
In this case, knowledge of the dilution is necessary only to determine
the corresponding value of $\sin(2\beta)$.
Using the value of the dilution
$D=0.166\pm0.018\,({\rm data})\pm0.013\,({\rm Monte\,Carlo})$
determined mainly with the $B^0$ mixing measurement discussed previously,
the value of $\sin(2\beta)$ is $1.8\pm1.1\,({\rm stat}.)\pm0.3\,({\rm syst}.)$.
The uncertainty on the dilution is contained in the systematic error
on $\sin(2\beta)$.
The non-physical value
is possible with low statistics; the corresponding limit is
$-0.2<\sin(2\beta)$ at $95\%$ C.L.

\begin{figure}[htb]
\postscript{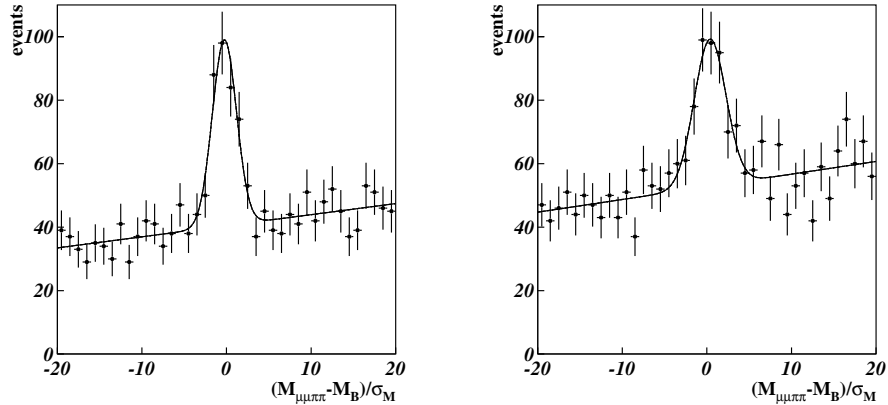}{0.8}
\caption{The normalized mass (see text) distribution of
$B^0/\bar{B}^0\rightarrow J/\psi K^0_{\rm S}$ candidates,
where $J/\psi\rightarrow\mu^+\mu^-$
and $K^0_{\rm S}\rightarrow\pi^+\pi^-$.
The left-hand (right-hand) plot is for candidates contained (not contained)
in the silicon microstrip detector.}
\label{fig:psikscand}
\end{figure}

As this first result was limited by statistics, the CDF collaboration
has continued to analyze the Run~I data set with the aim of improving
the statistical precision on $\sin(2\beta)$ as much as possible.
%
%
%
An updated measurement, which supersedes the published one
has been submitted recently for publication~\cite{ref:sin2betawww}.
The statistical power of the first published result has been
improved in two ways:
(1) the signal of 
$B^0/\bar{B}^0\rightarrow J/\psi K^0_{\rm S}$ has been doubled, and
(2) two opposite-side flavor tags (lepton tag and jet-charge tag)
have been added.
The published analysis used only candidates reconstructed in the
silicon microstrip detector, to ensure a precise measurement of
the $B$ proper decay time $t$.
Due to the large length (r.m.s.~of 30\,cm) of the distribution
of $p\bar{p}$ collisions along the beam line, only about $60\%$
of the interactions are contained the acceptance of the 50\,cm
long silicon detector.
The new analysis adds candidates that are not reconstructed in the
silicon microstrip detector.
The total signal of $B^0/\bar{B}^0\rightarrow J/\psi K^0_{\rm S}$
is $395\pm31$.
The normalized mass\footnote{
By normalized mass we mean the difference of the measured mass
from the world average $B^0$ mass divided by the estimated error
on the measured mass: [$M(\mu^+\mu^-\pi^+\pi^-)-M(B^0)$]$/\sigma_M$.}
distributions of the candidates reconstructed both inside and outside
of the acceptance of the silicon microstrip detector are shown in
Fig.~\ref{fig:psikscand}.
The mass resolution is approximately $\sigma_M\sim 10$\,MeV/$c^2$.
%
%
Candidates fully reconstructed in the silicon microstrip detector
have a proper decay time resolution of $\sigma_{ct}\approx 60\,\mu$m,
where as the candidates not fully contained in this detector
have $\sigma_{ct}\sim 300 - 900\,\mu$m.

Three methods of $b$~flavor tagging are applied to this sample:
the same-side tag described previously, a lepton tag and a jet-charge tag.
The lepton tag is identical to the tag used in a published
mixing analysis~\cite{ref:sltjctprd}.
The jet-charge tag is based on the jet-charge tag from the
published mixing analysis~\cite{ref:sltjctprd}, except that
the method of associating charged particles to the jet
was modified to increase the tag efficiency.
Any candidate track for the same-side tag is explicitly removed 
from the jet-charge determination.
If a lepton tag exists, then the jet-charge tag is ignored,
since the dilution of the lepton tag is much larger than
the dilution of the jet-charge tag.
These two requirements reduce the effect of correlations between
the tags and simplify the analysis.

The performance of the two opposite-side tags is quantified
using a signal of $998\pm51$ $B^{\pm}\rightarrow J/\psi K^{\pm}$ decays.
This data sample has similar kinematics to the
$B^0/\bar{B}^0\rightarrow J/\psi K^0_{\rm S}$ sample.
As mentioned previously, the same-side tag is expected to perform
differently for $B^+$ than for $B^0$, so this fully reconstructed
charged $B$ sample cannot be used to quantify the same-side tag.
Instead the performance is determined as described in the published
analysis~\cite{ref:sin2betaprl}.
The resulting flavor tag performance is summarized in Table~\ref{tab:tags}.
\begin{table}
\label{tab:tags}
\begin{tabular}{|l|c|c|c|}
\hline
Flavor tag & efficiency ($\epsilon$) & dilution ($D$)    & $\epsilon D^2$ \\
\hline
Lepton     & $(5.6\pm1.0)\%$         & $(62.5\pm14.6)\%$ & $(2.2\pm1.0)\%$ \\
Jet charge & $(40.2\pm2.2)\%$        & $(23.5\pm6.6)\%$  & $(2.2\pm1.3)\%$ \\
\hline
Same side (in Silicon) &
             $\sim70\%$               & $(16.6\pm2.2)\%$  & $(2.1\pm0.5)\%$ \\
Same side (not in Silicon) &
                                      & $(17.4\pm3.6)\%$ &               \\
\hline
\end{tabular}
\caption{The performance of the three flavor tags as applied
to the $B^0/\bar{B}^0\rightarrow J/\psi K^0_{\rm S}$ data sample.
For the same-side tag, $\epsilon$ and $\epsilon D^2$ are for
the total data sample (with and without silicon).
}
\end{table}
The performances of the individual tags are comparable;
the combined tagging effectiveness of all three flavor tags
is $\epsilon D^2 = (6.3\pm1.7)\%$.
The efficiency for tagging a $J/\psi K^0_{\rm S}$ with at least
one tag is $\sim 80\%$.

\begin{figure}[htb]
\postscript{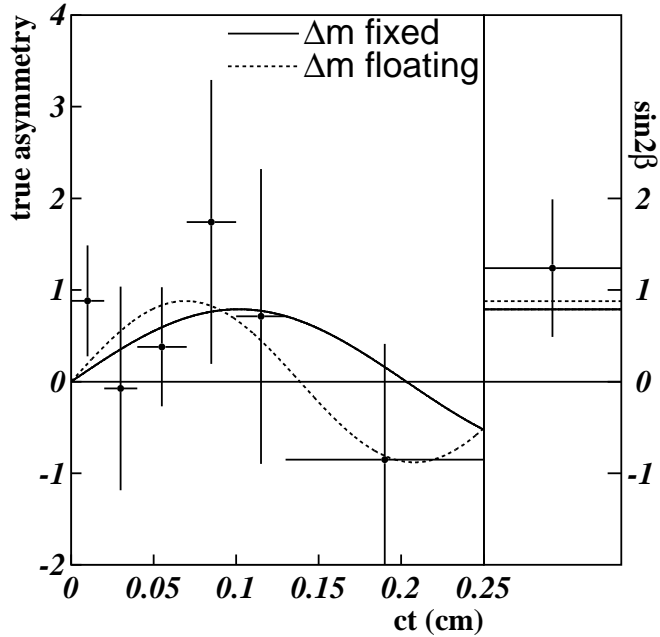}{0.6}
\caption{The updated determination of $\sin(2\beta)$.
The figure is described in the text.}
\label{fig:sin2betaprd}
\end{figure}

The value of $\sin(2\beta)$ is determined from the data using
the method of maximum likelihood.
The likelihood probability density includes the possibility
of detector biases ({\it e.g.}, differences in reconstruction
efficiency for positive and negative tracks) that may create
false asymmetries.
There are also terms that account for the behavior of backgrounds
to the signal.
These backgrounds are dominated by two sources:
(1) short-lived prompt background from prompt
$p\bar{p}\rightarrow J/\psi + X$ production with
a random $K^0_{\rm S}$, and 
(2) long-lived backgrounds from
$B\rightarrow J/\psi + X$ with
a random $K^0_{\rm S}$. 
A plot showing the data and the result of the fit is shown in
Fig.~\ref{fig:sin2betaprd}.
The result is $\sin(2\beta)=0.79^{+0.41}_{-0.44}$, where the
error includes contributions from both the statistics and systematic effects
(discussed below).
The left-hand side of the figure shows the data for which the candidates
are contained in the silicon microstrip detector; the points are the data
corrected by the dilution so that they represent the true asymmetry
(this is in contrast to Fig.~\ref{fig:sin2betaprl}).
The right-hand side of the figure shows the value of $\sin(2\beta)$
for the data not fully contained in the silicon microstrip detector.
The two curves are the results of the fit to both the left-hand and right-hand
data combined.
The amplitude of the solid curve is the value of $\sin(2\beta)$ quoted above.
The dashed curve is the same fit except that the value of $\Delta m_d$
is included as a fit parameter.
The result of this fit is
$\sin(2\beta)=0.88^{+0.44}_{-0.41}$ and
$\Delta m_d = 0.68\pm0.17\,{\rm ps}^{-1}$.
For comparison, a time-integrated determination of $\sin(2\beta)$
from these same data yields $\sin(2\beta)=0.71\pm0.63$;
the resulting error is about $50\%$ larger, which gives an
indication of the statistical improvement made possible by
making a time-dependent determination of $\sin(2\beta)$.

The statistics of the $J/\psi K^0_{\rm S}$ sample contribute $0.39$
to the total error.
The main systematic contribution to the error is $0.16$, which comes
from the uncertainty in the dilutions of the flavor tags.
This uncertainty is due to the limited statistics of
the data used to calibrate the performance of the tags.
As more data are accumulated in Run~II, the sizes of both the
$J/\psi K^0_{\rm S}$ sample and the calibration samples will increase
so that the statistics of the signal will continue to dominate
the error on $\sin(2\beta)$.

The variation of the negative logarithm of the likelihood with
$\sin(2\beta)$ follows a parabola near the minimum, so a confidence
limit on $\sin(2\beta)$ may be derived in a straightforward manner.
The Bayesian limit is $0<\sin(2\beta)<1$ at $95\%$ C.L.
If the true value of $\sin(2\beta)$ were zero, then the probability
of observing $\sin(2\beta)>0.79$ is $3.6\%$.
This result is the first compelling evidence that there is $CP$ violation
in $B$~hadron decays.

\section{Future Expectations}

At present, the CDF and D0 experiments are undergoing upgrades to improve
the performance of the detectors for the upcoming data taking period
that should begin in the year 2000.
The new accelerator component, the main injector, will increase the production
rate of antiprotons by approximately a factor of three over Run~I rates.
This increased rate will allow a substantial increase in instantaneous
luminosity.
The number of proton and antiproton bunches will be increased from
six each in Run~I, with bunches colliding every $3.5\,\mu$s, to thirty six each
in Run~II, with bunches colliding every $396$\,ns.
Eventually the machine will incorporate about 100 bunches each with
collisions every $132$\,ns.
The term ``upgrade'' is inappropriate as the changes to the detectors
are so substantial that much of the original components will be replaced.
For example, both CDF and D0 are replacing their entire charged-particle
tracking systems, and D0 is adding a 2\,T superconducting solenoid.
The electronics for components that were not upgraded have to be
replaced to accommodate the shorter time between bunch crossings.
Below we highlight the changes that are crucial for $B$~physics.

At CDF the important changes~\cite{ref:cdfup} are to the tracking system, the
trigger, and to particle identification capabilities.
There will be a new eight layer silicon system that extends
from a radius of $r=1.4$\,cm from the beam~line to $r=28$\,cm.
This system will include $r-\phi$ and $r-z$ information to allow
reconstruction in three dimensions.
It will cover out to a pseudorapidity of $|\eta|<2$, which will
double the acceptance as compared to the Run~I detector.
A new central drift chamber has been constructed.
The drift cell size has been reduced by a factor of four,
and ultimately with use of a fast gas, the drift time will
be reduced by a factor of eight from Run~I.
The goal is to maintain the excellent reconstruction efficiency
and momentum resolution of the Run~I drift chamber.
The trigger system is pipelined to allow ``deadtimeless'' operation.
In particular, fast tracking is now possible at level~one, and
information from the silicon detectors will be available at level~two,
making it possible to trigger on the presence of tracks originating
from the decay of long-lived $B$~hadrons.
This should allow triggering on all hadronic decays of $B$ hadrons
such as $B^0_s\rightarrow\pi^+D_s^-$, with $D_s^-\rightarrow\phi\pi^-$,
which is important for $B^0_s$ mixing,
and $B^0/\bar{B}^0\rightarrow\pi^+\pi^-$,
which is important for $CP$ violation.
Finally a time-of-flight detector will make kaon identification possible
in a momentum range that is especially interesting for
$b$~flavor identification.

The D0 detector is undergoing an even more
radical change~\cite{ref:dzeroup} than CDF.
A superconducting solenoid (2\,T) has been installed
to allow momentum measurements.
The main tracking device will be an eight-layer scintillating fiber
detector with full coverage out to $|\eta|<1.7$.
A silicon microstrip tracker will be installed.
It will consist of six barrel detectors, each with four layers,
with both $r-\phi$ and $r-z$ read out, and sixteen disks extending
out to $|z|<1.2$\,m along the beam line.
With this new spectrometer and tracking system,
D0 will tag $B$ decays using displaced vertices.
Finally, with improvements to the muon system and trigger,
as well as the existing liquid argon calorimeter, D0 should
be competitive with CDF for $B$~physics in Run~II.

CDF and D0 will address many important questions in $B$~physics in Run~II.
Many of the relevant measurements have already been investigated using
the Run~I data.
Among the more important goals are
(1)
the precise measurement of $|V_{\rm td}/V_{\rm ts}|$ from
$B^0_s-\bar{B}^0_s$ flavor oscillations {\boldmath ($\dagger$)} 
(or from a measurement of $\Delta\Gamma_s/\Gamma_s$)
or from  radiative decays, {\it e.~g.}, the rate of
$B^0_s\rightarrow K^{*0}\gamma$
compared to $B^0_s\rightarrow \phi\gamma$~{\boldmath($\dagger$)};
(2)
the observation of $CP$~violation in
$B^0/\bar{B}^0\rightarrow J/\psi K^0_{\rm S}$ and the
precise measurement of $\sin(2\beta)$;
(3)
the observation of $CP$~violation in $B^0/\bar{B}^0\rightarrow\pi^+\pi^-$
and a precise measurement of the $CP$~asymmetry,
which is related to $\sin(2\alpha)$;
(4)
the search for large $CP$~violation in
$B^0_s/\bar{B}^0_s\rightarrow J/\psi\,\phi$ {\boldmath ($\dagger$)};
(a large asymmetry would be
an unambiguous signal of physics beyond the Standard Model);
(5)
the observation of decay modes related to angle $\gamma$:
$B^0_s\rightarrow D^{\pm}_s K^{\mp}$~{\boldmath ($\dagger$)}
and $B^+\rightarrow \bar{D}^0 K^+$;
(6)
the observation of rare decays such as
$B^+\rightarrow\mu^+\mu^- K^+$, $B^0\rightarrow\mu^+\mu^- K^{*0}$, and
$B^0_s\rightarrow\mu^+\mu^- \phi$~{\boldmath ($\dagger$)};
(7)
the study of the $B^+_c$~meson {\boldmath ($\dagger$)}
and $b$ baryons {\boldmath ($\dagger$)} .
As long as the $B$~factories remain on the $\Upsilon(4S)$ resonance,
the topics marked by a dagger~{\boldmath ($\dagger$)} will be unique to
hadron colliders.

Many studies of these future topics have been performed.
Here we summarize expectations for two more important measurements:
(1) $\sin(2\beta)$ and (2) $\Delta m_s$.
A more complete discussion can be found in~\cite{ref:paulini}.
These projections ({\it e.g.}, flavor tag performance) are based on Run~I
data from CDF as much as possible, and assume an integrated luminosity
of 2~fb$^{-1}$ collected during the first two years of operation.
Although the expectations are specific to CDF, similar sensitivity
should be possible with the D0 detector.

For the measurement of $\sin(2\beta)$, the signal size should
increase to 10\,000 $B^0/\bar{B}^0\rightarrow J/\psi K^0_{\rm S}$ decays,
where $J/\psi\rightarrow\mu^+\mu^-$ and $K^0_{\rm S}\rightarrow\pi^+\pi^-$.
The $b$ flavor tags will be calibrated with samples of
40\,000 $B^{+}\rightarrow J/\psi K^{+}$ (and charge conjugate) decays and
20\,000 $B^0\rightarrow J/\psi K^{*0}$ (and charge conjugate) decays.
The expected combined flavor tag effectiveness of the same-side,
lepton, and jet-charge tags discussed earlier is $\epsilon D^2=6.7\%$.
The resulting estimate of the error on $\sin(2\beta)$
is  $\delta(\sin(2\beta))=0.084$.
This uncertainty includes the systematic uncertainties in the dilutions
due to the statistics of the calibration samples.
The time-of-flight detector will make it possible to use a new
opposite-side flavor tag based on kaons: the decays of
$B$ hadrons containing $\bar{b}$ ($b$) quarks usually produce $K^+$ ($K^-$).
With this additional flavor tag, the total flavor tag effectiveness
could increase to $\epsilon D^2=9.1\%$~\cite{ref:prop909}.

Based on current experimental measurements and theoretical predictions,
it is possible to predict the value of $\sin(2\beta)$.
For example, S.~Mele~\cite{ref:mele} predicts $\sin(2\beta)=0.75\pm0.09$
(not including the latest measurement of $\sin(2\beta)$ from CDF).
This current projection rivals the expected precision
of the measurement in Run~II, therefore, to really test our predictions
and consistency within the Standard Model, even more precise measurements
of $\sin(2\beta)$ will be necessary.
This motivates the need for future experiments such as BTeV~\cite{ref:btev}
at the Tevatron.

The other crucial measurement, which is attainable only at hadron colliders,
is the measurement of $\Delta m_s$.
The upgraded CDF trigger will make it possible to trigger on fully
reconstructed hadronic $B$ decays.
A large sample of 20\,000 fully reconstructed $B^0_s$ decays in
the modes $B^0_s\rightarrow D_s^- \pi^+, D_s^- \pi^+\pi^-\pi^+$,
with $D_s^-\rightarrow \phi\pi^-, \bar{K}^{*0}K^-$ is expected.
The time-of-flight detector will double the expected $b$~flavor tagging
efficiency to $\epsilon D^2=11.3\%$ by exploiting kaon identification
in the same-side tag of $B^0_s$ mesons (see Fig.~\ref{fig:sst}).
The CDF silicon system will have excellent impact parameter resolution,
and the expected proper lifetime resolution is $45$\,fs.
With all these improvements combined, a better than
five-sigma measurement of $\Delta m_s$ is possible
for $\Delta m_s < 40$\,ps$^{-1}$~\cite{ref:prop909}.

\section{Summary}

In Run~I many important measurements of $B$ hadron decay properties were
made at the Tevatron, despite severe trigger restrictions.
Many of these measurements are similar or better in precision to
measurements from $e^+e^-$ colliders; some are unique to the Tevatron.
In particular, a first indication of $CP$ violation in the $B$ system
was found yielding $\sin(2\beta)=0.79^{+0.41}_{-0.44}$.
%

In Run~II, improved detectors will increase the scope of $B$ physics
at the Tevatron. 
The precision on $\sin(2\beta)$ should be competitive (and complementary)
with measurements at the $B$ factories.
Studies of $CP$ violation in $B^0_s$ decays and the measurement of
$\Delta m_s$ will be unique to hadron colliders.
The Tevatron will play a crucial, unique role in our test of the
CKM matrix.

\section{Acknowledgements}

I would like to thank the organizers of this conference for an enjoyable
informative, and well organized meeting.
This work is supported by DOE grant number DE-FGO2-95ER40893.

\end{document}